\documentclass[10pt]{IEEEtran}
\usepackage{cite}
\usepackage{amsmath,amssymb,amsfonts}
\usepackage{graphicx}
\usepackage{textcomp,nicefrac}
\def\BibTeX{{\rm B\kern-.05em{\sc i\kern-.025em b}\kern-.08em
T\kern-.1667em\lower.7ex\hbox{E}\kern-.125emX}}
\begin{document}
\title{Serenity-S1, a CMS Tracker readout and data processing card for HL-LHC}
\author{G. Fedi, L.E. Ardila-Perez, M. Balzer, M. Fuchs, M. Holmberg, A. Howard, G.M. Iles, H.A. Krause, T. Mehner, D. Parker, M. Pesaresi, A. Rose, K. Whalen, T. Williams, and J. Zhao on behalf of the CMS Collaboration

\thanks{G. Fedi (e-mail: giacomo.fedi@cern.ch), A. Howard, G. Iles,  D. Parker, M. Pesaresi, and A. Rose are with the Imperial College London,  London,  SW7 2AZ UK. L.E. Ardila-Perez, M. Balzer, M. Fuchs, H. Krause, and T. Mehner are with Karlsruhe Institute of Technology, Hermann-von-Helmholtz-Platz 1, D-76344 Eggenstein-Leopoldshafen, Germany. T. Williams and K. Whalen are with STFC Rutherford Appleton Laboratory, Harwell Campus, Didcot, OX11 OQX, UK. M. Holmberg is with  University of Bristol, Queens Road, Bristol, BS8 1QU, UK. J. Zhao is with Chinese Academy of Sciences 19B YuQuan Road 100049 Beijing. H. Krause and T. Mehner acknowledge the support by the Doctoral School “Karlsruhe School of Elementary and Astroparticle Physics: Science and Technology”.}}

\maketitle

\begin{abstract}
In the coming years, the CERN Large Hadron Collider will be upgraded, increasing the rate of particle collisions. A key component of this upgrade is the new Tracker detector, composed of silicon sensors that will generate data at the order of 10 terabytes per second. This massive data stream must be efficiently routed to processing units and filtered before being stored on disk. 
 
Serenity-S1 features an AMD VU13P UltraScale+ FPGA, utilizing all 128 high-speed serial transceivers. Most of these (120) are connected via Samtec 12 channel FireFly connectors, enabling data transmission over optical fibers at 25 Gb/s per channel. The same connector is populated with a different Samtec FireFly to communicate with CMS Tracker at up to 10 Gb/s per channel. 

The ability to populate the Samtec Firefly connector with different optics depending on the task required makes the card extremely flexible. This aspect coupled with the large FPGA has allowed Serenity-S1 to be used for many other applications both inside and outside the CMS experiment. The card is built on the ATCA form factor which provides the 400 W of power and cooling required and is well suited to the existing infrastructure in the CMS Service Cavern, which is located 100 m underground, adjacent to the CMS detector. Critical board management tasks and the IPMI interface are supplied by the OpenIPMC project in the form of a low profile DIMM. 

An AMD Kria K26 SoM running Alma 9 Linux performs the remaining board management via software that mimics the board hardware, allowing autonomous routing of commands. The SoM also performs board application control via a firmware and software suite that provides all the core infrastructure, allowing users to focus on their specific application. An application of the Serenity-S1 board is demonstrated in the CMS Tracker project, where several boards have already been deployed in integration and R\&D activities.
\end{abstract}


%%%%%%%%%%%%%%%%%%%%%%%%%%%%%%%%%%%%%%%%%%%%%%%%%%%%%%%%%%%%%%%%%%%%%%
\section{Introduction}
The Compact Muon Solenoid (CMS)~\cite{CMS:2008xjf} detector at CERN’s Large Hadron Collider is designed to address a wide spectrum of high-energy-physics questions, ranging from precision studies of the Higgs boson to searches for extra spatial dimensions and dark-matter particles. During the next decade the accelerator will be converted into the High-Luminosity LHC (HL-LHC), increasing the instantaneous collision rate by roughly an order of magnitude. To operate under these more demanding conditions CMS must be equipped with radiation-hard sensors and electronics; finer detector granularity that can disentangle up to about 200 overlapping proton–proton interactions every 25 ns; substantially higher read-out bandwidth, and a first-level trigger that can sustain rates near 750 kHz without compromising physics reach.

A complete replacement of the silicon Tracker is central to this program. The new system will include roughly 17,000 radiation-tolerant silicon modules arranged in four inner pixel layers and six outer strip/macro-pixel layers. Charged-particle trajectories will be measured with about 10 µm precision, and the 3.8 T solenoidal field will yield accurate momentum determination over a larger rapidity range than today’s detector. Hit data flow out of the Tracker along optical fibres operating at up to 10 Gb/s per link, producing an aggregate raw rate of order 10 TB/s.

\begin{figure}[t]
\centerline{\includegraphics[width=2.7in]{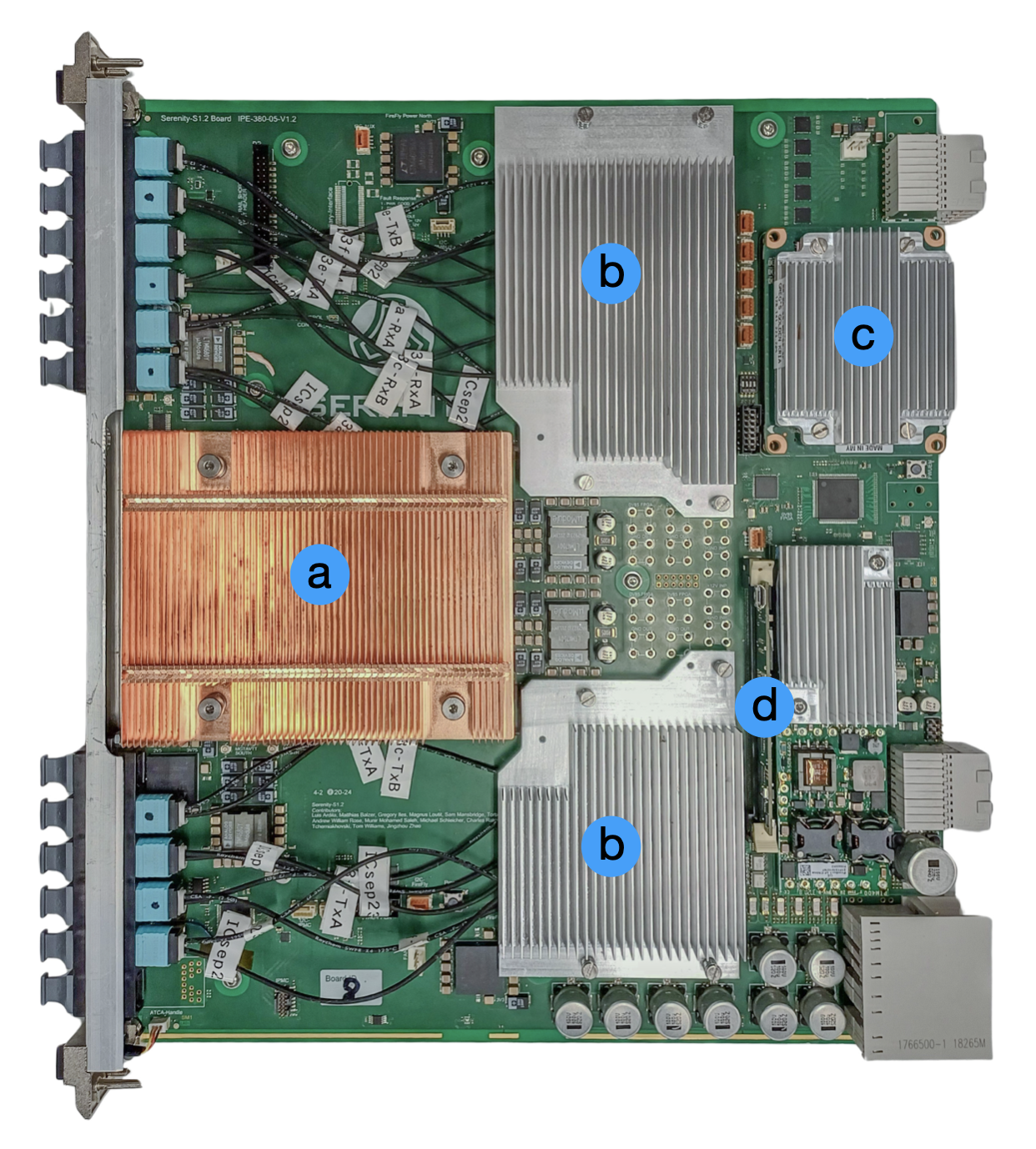}}
\caption{View of a Serenity-S1 board: \raisebox{.5pt}{\textcircled{\raisebox{-.9pt} {a}}}  VU13P FPGA below a copper vapour chamber heat-sink; \raisebox{.5pt}{\textcircled{\raisebox{-.9pt} {b}}} 21 Samtec FireFly sockets below two aluminium heat-sinks; \raisebox{.5pt}{\textcircled{\raisebox{-.9pt} {c}}} Kria K26 SoM; \raisebox{.5pt}{\textcircled{\raisebox{-.9pt} {d}}} OpenIPMC DIMM mezzanine.}
\label{fig1}
\end{figure}

The CMS backend system will implement 721 Serenity-S1 boards. Each board hosts an AMD/Xilinx Virtex UltraScale+ VU13P FPGA and 21 Samtec FireFly optical transceivers, together providing more than 3 Tb/s of bidirectional bandwidth. Firmware running on the FPGA performs real-time decoding, formatting, and routing of detector information to the downstream trigger and data-acquisition stages. Because Serenity-S1 is fully reconfigurable, the same hardware platform can serve multiple CMS subsystems, simplifying maintenance and ensuring the flexibility required for a decade of HL-LHC operation. 
%%%%%%%%%%%%%%%%%%%%%%%%%%%%%%%%%%%%%%%%%%%%%%%%%%%%%%%%%%%%%%%%%%%%%%

\section{Hardware Architecture}
\subsection{Concept and Cooling}
A key objective of the board design was to ensure that the optics could be easily mounted, with simple routing of the on-board fibre to the front panel, whilst at the same time ensuring that they remained thermally decoupled from the FPGA. This resulted in the board being split into 3 sections, each with their own cooling air column (i.e. FPGA, optics  and service regions).  This approach has kept the temperature of the optics below 50$^\circ$C, ensuring that they have a long service life.  A heat-sink with integrated vapour chamber dissipates up to 200 W from the FPGA whilst remaining within the narrow envelope of the ATCA card.

\subsection{System on Module and Board Controller}
A Kria K26 system-on-module (SoM), featuring a quad-core Cortex-A53 clocked at 1.2 GHz, handles board-level orchestration: it boots an Alma Linux 9 distribution, launches user applications and continuously supervises system health, all reachable over standard Gigabit-Ethernet.

Low-level management is provided by an OpenIPMC Mini-DIMM mezzanine~\cite{Calligaris:2023nab}. This open-hardware card uses an STM32H745 dual-core microcontroller running the OpenIPMC firmware, giving the ATCA blade a vendor-neutral, easily customisable alternative to proprietary IPMCs. The firmware has been tailored for this board to meet every ATCA requirement—power sequencing, sensor read-back and hot-swap control—and exposes remote services such as IPMI, Xilinx Virtual Cable (XVC) and Serial-over-LAN for seamless debugging and maintenance.

\subsection{Additional I/O and Clock Network}
% Tri-stack PCB structure
An optional Rear Transition Module can be plugged in to the back of the card, providing an electrical data path to/from the FPGA.  It also allows external clocks to be routed into the clock network, which is based on 6 jitter-cleaning Microchip ZL30274s.  Clock signals can also originate from the ATCA backplane, be internally generated, or be recovered from incoming optical fibres.

%%%%%%%%%%%%%%%%%%%%%%%%%%%%%%%%%%%%%%%%%%%%%%%%%%%%%%%%%%%%%%%%%%%%%%
\section{Firmware and Software Ecosystem}
EMP is a turnkey firmware and software framework: it supplies the control bus master, fixed-latency timing, and other high-speed serial links, plus playback/capture buffers in a common top-level design, so users need only plug in their own payload logic. A configurable build system then auto-selects the right device-specific files for any supported FPGA package, avoiding code duplication and keeping the build workflow uniform across boards. The framework is also composed of a software counterpart running on the Kria SoM for the interaction with the firmware. 

Hardware configuration is managed by SMASH, a plug-in control framework that represents the board as a hierarchical tree of independent elements. By mimicking the board's hardware in software, SMASH enables autonomous command routing and efficient board management.

%%%%%%%%%%%%%%%%%%%%%%%%%%%%%%%%%%%%%%%%%%%%%%%%%%%%%%%%%%%%%%%%%%%%%%
\section{Serenity-S1 in the Outer Tracker}

The Outer Tracker---the strip and strip/macro-pixel region of the CMS detector---produces two independent data streams from each module: low-latency trigger information and full-resolution hit data. Serenity-S1 ATCA boards receive and process these streams. The onboard FPGAs reformat and direct the data accordingly: every 25 ns an order of 10k hit segments are sent to pattern-recognition engines, which reconstruct order of 100 tracks within 4~$\mu$s. Meanwhile, the full-resolution data are routed to concentrator boards that aggregate all channels before forwarding them to the event builder.

The read-out plant is divided into nine azimuthal sectors, each serviced by 24 Serenity-S1 boards; the Outer-Tracker data acquisition will host roughly 260 boards, including spares.

Real-time routing is executed in the FPGA, while the on-board Kria SoM supervises configuration, slow-control and health monitoring of both firmware and the silicon modules.  Because the SoM’s resources are modest, most control software is built off-board and deployed in Docker containers for validation on the target hardware.

%%%%%%%%%%%%%%%%%%%%%%%%%%%%%%%%%%%%%%%%%%%%%%%%%%%%%%%%%%%%%%%%%%%%%%
\section{Test Results and Production Status}

A pilot run confirmed the electrical and signal-integrity goals: every high-speed channel had at least a 30\% opening in its bath-tub at a BER~$< 10^{-12}$; a prerequisite for ten years of continuous HL-LHC operation.

Dozens of production Serenity-S1 boards have since been assembled.  Each card passes through an automated factory test suite.  On-board debug headers expose all power rails, enabling rapid shorts/opens checks, while scripting software interrogates every device.  The test sequence exercises clock trees, verifies I\,$^2$C inventories, and stresses all multi-gigabit transceivers via local copper loop-backs.  Boards that clear the suite proceed directly to system integration.

%%%%%%%%%%%%%%%%%%%%%%%%%%%%%%%%%%%%%%%%%%%%%%%%%%%%%%%%%%%%%%%%%%%%%%
\section{Conclusion}
Serenity‑S1 demonstrates that a single‑blade ATCA solution can satisfy the multi‑terabit‑per‑second, low‑latency demands of the HL‑LHC Tracker while offering firmware and software agility for future upgrades. Its modular optical strategy and open management layer position it as a reusable platform across several CMS subsystems and potentially other experiments.

\end{document}